\documentclass[%
 reprint,
 amsmath,amssymb,
 aps,
prx
]{revtex4-2}

\usepackage{graphicx}
\usepackage{dcolumn}
\usepackage{bm}
\usepackage{amsmath}
\usepackage{amssymb}

\newcommand{\cmt}[1]{{}}

\usepackage{xcolor}

\begin{document}
\preprint{APS/123-QED}

\title{Junction-free microwave two-mode radiation from a kinetic inductance nanowire}


\author{Yufeng Wu}
\author{Mingrui Xu}
\author{Hong X. Tang}%
 \email{hong.tang@yale.edu}
\affiliation{%
 Department of Electrical Engineering, Yale University, New Haven, Connecticut 06520, USA
}%

\date{\today}

\begin{abstract}
Parametric down-conversion is a widely exploited technique in optics to produce entangled states of photons for quantum information processing and quantum sensing. In the microwave domain, devices based on Josephson junctions, such as Josephson parametric amplifiers (JPAs) and voltage-biased Josephson junctions, have been successfully utilized to generate such states. However, their high susceptibility to magnetic fields has posed challenges in many applications. Here we demonstrate the generation of two-mode squeezed states via four-wave-mixing in a superconducting nanowire resonator patterned from NbN. The NbN nanowire exhibits a strong Kerr nonlinearity, resulting in the emission of a signal-idler pair with a cross-correlation of $g^{(2)}(0) = 11.9$. Owing to the magnetic resilience and high critical temperature ($T_c$) of NbN, our microwave parametric sources based on kinetic inductance promise an expanded range of potential applications. 

\end{abstract}

\maketitle

\section{Introduction}

Nonclassical states are quantum states characterized by photon statistics that do not follow classical descriptions. Among these states, two-mode non-classical states are particularly intriguing due to their strong quantum correlation. Two-mode squeezed states \cite{eichler2011observation}, in particular, play a vital role in a wide range of applications, such as quantum teleportation \cite{furusawa1998unconditional, milburn1999quantum}, quantum sensing \cite{zhuang2018distributed, lawrie2019quantum, backes2021quantum, wurtz2021cavity, bienfait2017magnetic, didier2015heisenberg, liu2022noise} and continuous-variable quantum computing \cite{lloyd1999quantum, gu2009quantum}. In the region where the average photon number is very small, such states can be approximated as photon-pair states, which is an indispensable component in Bell inequality testing \cite{tittel1998violation, weihs1998violation, storz2023loophole}, photonic quantum computing \cite{zhong2020quantum} and global quantum networks \cite{kimble2008quantum, simon2017towards, wehner2018quantum}. 

The generation of such non-classical two-mode radiation in the microwave domain has predominantly relied on Josephson Parametric Amplifiers (JPAs) \cite{eichler2011observation, bergeal2012two, flurin2012generating, barzanjeh2020microwave}. These devices utilize parametric down-conversion to transform pump photons into pairs of signal and idler photons exhibiting strong correlation. Recently, this process has also been demonstrated in DC-biased Josephson junctions through inelastic Cooper pair tunneling \cite{westig2017emission, peugeot2021generating}, traveling wave parametric amplifiers (TWPAs) \cite{esposito2022observation, qiu2023broadband}, and access higher energy level of a superconducting transmon qubit \cite{gasparinetti2017correlations}. However, without exception, all these devices are based on Josephson junctions, which are sensitive to external flux. Therefore, they are not compatible with quantum systems that require strong magnetic fields, for example, two-mode squeezed state enhanced dark-matter axion search \cite{wurtz2021cavity, jiang2023accelerated}. The traditional approach to mitigating the influence of the magnetic field is to move the parametric amplifier away from the magnet and protect it with layers of shield \cite{backes2021quantum}. In some scenarios, flux feedback is required to cancel the excessive magnetic field \cite{brubaker2017first}. On the other hand, higher-order nonlinearities in JJ-based devices have been identified as a limiting factor for the amplification and squeezing \cite{boutin2017effect}. These limitations have prompted researchers to explore alternative approaches to overcome these challenges and expand the potential applications of these devices in various quantum systems.

Niobium nitride (NbN) and niobium titanium nitride (NbTiN) have been gathering attention due to their high kinetic inductance nonlinearity properties. Utilizing a nanowire structure, these materials can function as parametric amplifiers with quantum-limited added noise \cite{parker2022degenerate, xu2023magnetic}. Furthermore, NbN exhibits a high critical temperature (around 10 K), facilitating kinetic inductance traveling-wave parametric amplifiers at 4 Kelvin with an added noise temperature of 6.5 K \cite{malnou2022performance}. Additionally, it demonstrates robust in-plane magnetic field resilience, allowing NbN based resonators to operate at magnetic field up to 6 T \cite{yu2021magnetic}, and quantum amplifiers to operate up to 2 T \cite{xu2023magnetic, khalifa2023nonlinearity}.  

In this work, we observed non-classical two-mode radiation from a kinetic inductance nanowire based on NbN. We send two symmetric detuned pumps to the device to drive the parametric down-conversion, and demodulate the two itinerant modes that are evenly detuned from the resonance frequency. The non-classicality is characterized by a violation of the classical bound of the second-order cross-correlation. Our investigation reveals an second-order cross-correlation reaching $11.91 \pm 0.68$ with MHz photon-pair generation rate. These results solidify the benefits of kinetic inductance devices as potential replacements for JJ-based parametric amplifiers. 

\section{Hamiltonian and Photon Correlations}
\begin{figure}[t]
    \centering
    \includegraphics[width=\linewidth]{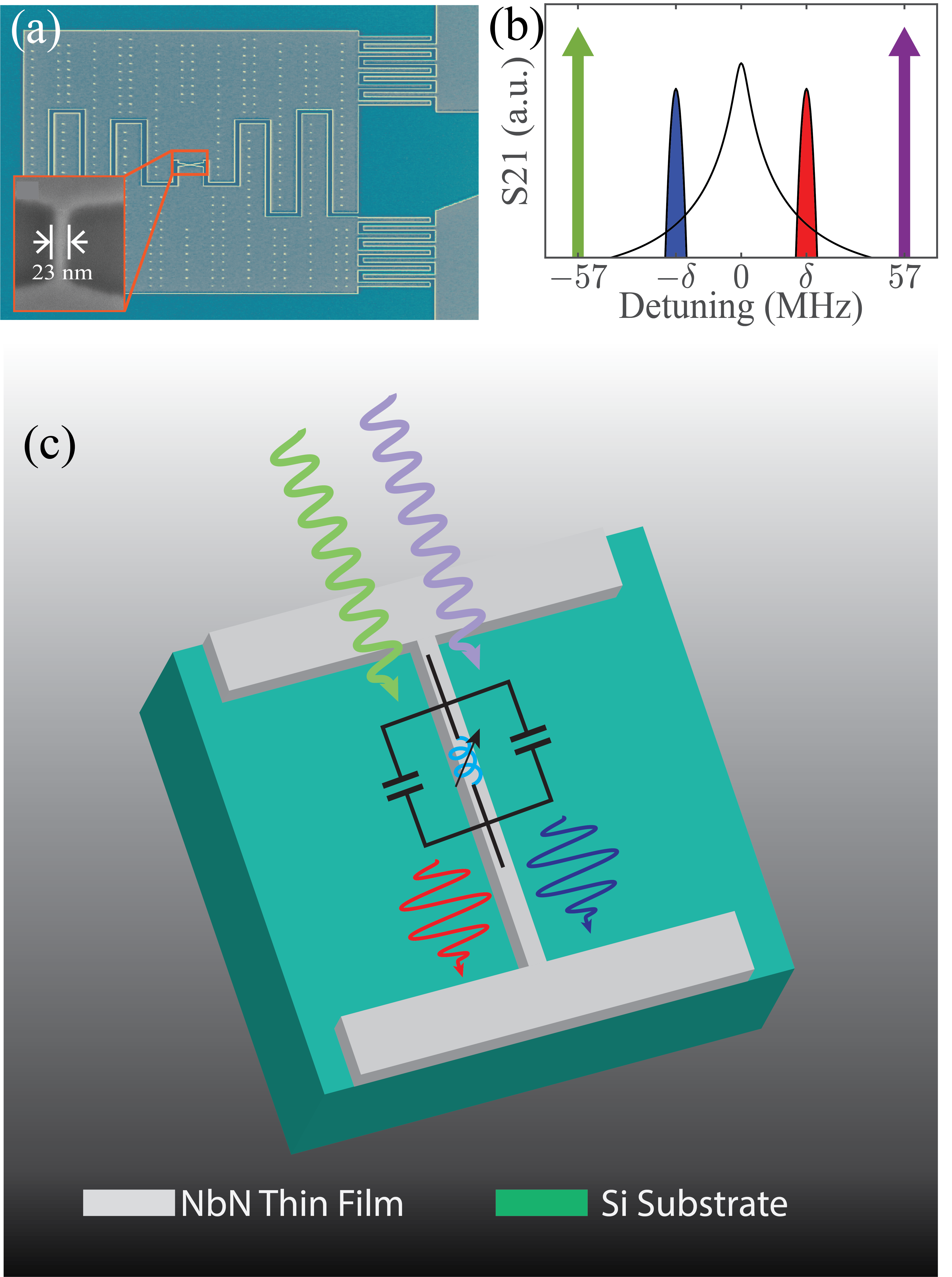}
    \caption{Device and illustration of the two-mode generation. (a) Optical image of the NKPA. The brighter area represents NbN, while the darker area signifies the silicon substrate. The inset SEM image displays a zoomed-in view of the nanowire. (b) Pumping schematic of the NKPA device. The two arrows illustrate the symmetric pump, and the two shaded regions represent the signal and idler modes to be demodulated. (c) Artistic illustration of the two-mode generation from an NbN nanowire. The inserted is the equivalent lumped element circuit diagram of the NKPA.} 
    \label{fig:cartoon}
\end{figure}
The nanowire kinetic-inductance parametric amplifier (NKPA) device is fabricated from NbN with a thickness of 4 nm deposited on silicon substrate via atomic-layer deposition (ALD) \cite{xu2023magnetic}. The device is designed as an LC resonator that incorporates highly nonlinear kinetic inductance from the nanowire, shown in Fig.\ref{fig:cartoon} (a). Under a symmetric two-tone drive, as shown in Fig. \ref{fig:cartoon} (b), the system Hamiltonian in the interacting picture becomes (ignoring the self-Kerr and cross-Kerr component) \cite{xu2023magnetic} 
\begin{equation}
    \begin{aligned}
H_{\text {int}} = \hbar \left(\epsilon a^{\dagger}b^{\dagger}+ \epsilon^{*} ab\right),
\end{aligned}
\end{equation}    
where $\epsilon = KA_1A_2e^{i (\phi_1 + \phi_2)}$ is the parametric down-conversion coefficient, and $A_1, A_2$, $\phi_1, \phi_2$ are the two-pump amplitudes and phases, $K = 2\pi \times 110$ kHz is the single-photon Kerr nonlinearity. Detailed derivation can be found in Appendix \ref{app:Hamiltonian}. This represents a two-mode squeezing Hamiltonian. Non-classical correlations emerge for two modes symmetrically positioned around the resonance frequency. The relative position and total momentum between the two modes are squeezed below the standard vacuum limit \cite{eichler2011observation}. 
However, due to the added noise from the amplifier in the subsequent circuit, this can only be observed when operating the amplifier in the high-gain region, and may sometimes necessitate an additional quantum-limited preamplifier \cite{flurin2012generating}. In this work, we operate the amplifier in the low-gain region, the system serves as a photon-pair source, as illustrated in Fig. \ref{fig:cartoon} (c). The non-classical nature of the two-mode squeezing state is characterized by the second-order correlation

\begin{equation}
    \label{equ:g2_cross}
    g^{(2)} _{\alpha\beta} (\tau) = \frac{\langle \beta^{\dagger} (0)\alpha^{\dagger}(\tau)\alpha(\tau)\beta(0) \rangle )}{\langle \alpha^{\dagger}(\tau) \alpha(\tau) \rangle \langle \beta^{\dagger}(0) \beta(0) \rangle},
\end{equation}
with $\alpha, \beta \in \{a, b\}$. In analogy with optics, the numerator represents the coincidence count rate and the denominator represents the product of the count rate of the two respective modes. It is important to note that while these values depend on the system loss, their quotient $g^{(2)}(\tau)$ remains independent of system loss \cite{guo2017parametric}.  For a classical state, zero-delay cross-second order correlation is smaller than the average of the two auto-second order correlation 
\begin{equation}
    \label{equ:inequality}
    g^{(2)}_{ab}(0) \leq \frac{g^{(2)}_{aa}(0)+ g^{(2)}_{bb}(0)}{2}.
\end{equation}
A two-mode squeezed state can be expressed as \cite{lvovsky2015squeezed}
\begin{equation}
    |\Phi_{TM}\rangle =  \frac{1}{\cosh{r}} \sum_n (\tanh{r})^n |n, n\rangle,
    \label{equ:tms_sq_fock_basis}
\end{equation}
where $r=\epsilon \tau$ is the squeezing parameter, $\tau$ is the cavity decay time. Under a weak drive, where $r$ is small, the two-mode squeezed state is reduced to a mix of vacuum and photon-pair states. Therefore, a higher-than-2 $g^2_{ab}(0)$ can be observed and thus violate Eq. (\ref{equ:inequality}), inferring the non-classical characteristics. 
\section{Experimental Setup}
\begin{figure}[t]
    \centering
    \includegraphics[width=\linewidth]{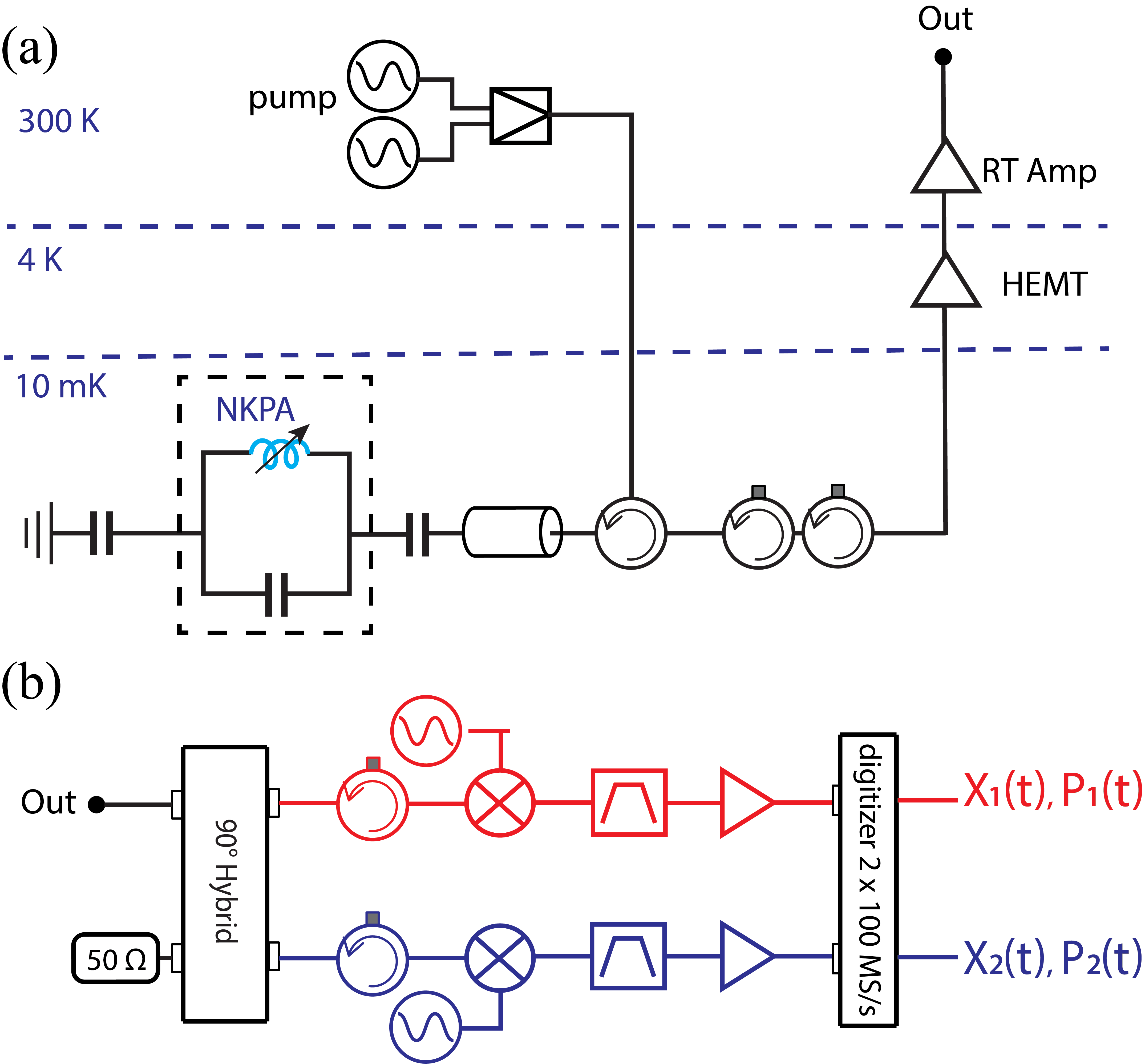}
    \caption{Schematic of the experimental setup. (a)  Cryogenic electrical circuit connection schematic. The two pumps are combined and directed toward the NKPA device located in the MXC chamber. (b) The room-temperature measurement setup. The output signal is split into two demodulation channels with a $90^{\circ}$ hybrid. The intermediate-frequency filter is centered at $74.6$ MHz with a bandwidth of $3$ MHz. } 
    \label{fig:setup}
\end{figure}
The cryogenic circuit connection schematic is shown in Fig. \ref{fig:setup} (a). The NKPA device was placed in the mixing chamber of a dilution refrigerator and cooled to a temperature of 10 mK. To send the two pump tones to the device, they were symmetrically detuned from the resonance frequency of $7.359$ GHz by $57$ MHz and subsequently combined.  In order to prevent the impact of room-temperature thermal noise and pump phase noise, the pump tone is attenuated 60 dB to the device. The output signal is amplified by a high electron mobility transistor (HEMT) amplifier and room-temperature (RT) amplifier. 
 
Figure \ref{fig:setup} (b) depicts the room-temperature measurement setup. The output signal is directed towards a $90^{\circ}$ hybrid coupler, where one of the input ports is terminated by a 50 $\Omega$ resistor, dividing the signal into two channels. The microwaves are down-converted to $75$ MHz using a mixer, with local oscillator frequencies $\omega_{\mathrm{LO1,2}} = f_c \pm (75\, \mathrm{ MHz} + \delta$), where $\delta = 2$ MHz represents the detuning of the signal/idler tone from the resonance center. The down-converted signal and idler tone are processed through bandpass filters with a center frequency of $74.60$ MHz and a bandwidth of $3.88$ MHz. Following intermediate frequency amplification, the signal and idler tones are transmitted to two detection channels of the digitizer operating at a sampling rate of 100 MSp/s. This aliases both the signal and idler frequency to an effective frequency of $25$ MHz, while the signal-to-noise ratio (SNR) is not degraded due to the narrow-bandwidth bandpass filter. The digitizer converts the analog signal to digital voltage with an 8-bit precision and sends it to a graphics processing unit (GPU) for real-time data streaming and processing. To further reduce the spectral overlap between the signal and idler and increase the mode selectivity, a digital filter with a bandwidth of 0.86 MHz is applied to the data in both channels, after which the second-order auto- and cross-correlation is calculated and the result is transmitted to the CPU. 

Notably, the high noise temperature of the HEMT amplifier causes the correlation to be obscured by the amplified noise, making it challenging to extract directly. To overcome this issue, we employed the post-processing method of ``ON/OFF" measurement, which involves taking measurements under two different conditions - the ``ON" and ``OFF" states \cite{grimm2015josephson, westig2017emission, peugeot2021generating}. Specifically, the ``ON" measurement, represents the statistics of both the signal and noise, while the ``OFF" measurement, with one of the pump shift 80 MHz away to destroy the frequency-matching condition, only characterizes the statistics of the noise. By combining both sets of measurements, we could extract the statistics corresponding to the signal alone. The detailed discussion can be found in Appendix \ref{sec:dataprocessing}.  

\section{Result}

\begin{figure}
    \centering
    \includegraphics[width=\linewidth]{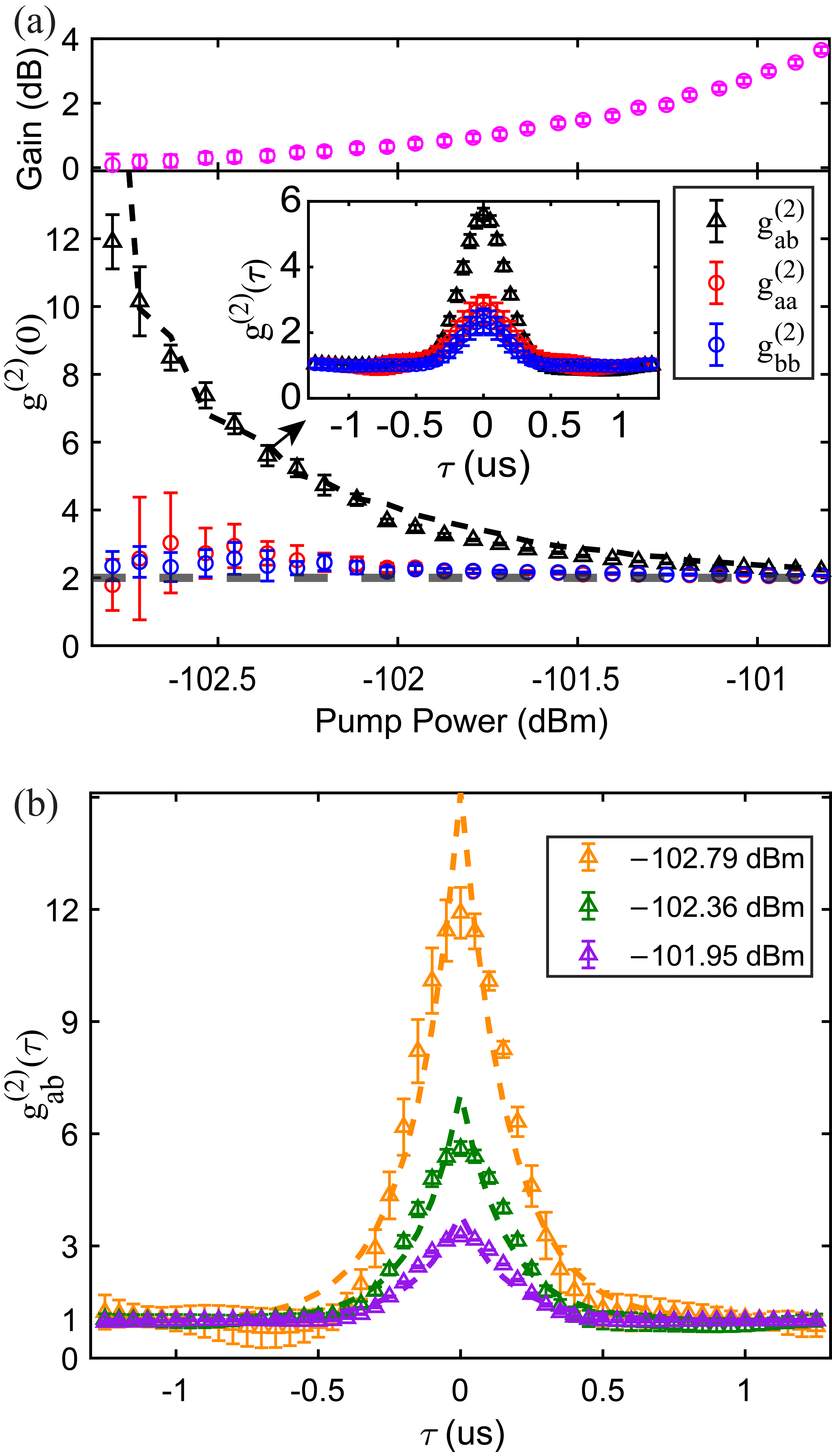}
    \caption{(color online) (a) Parametric amplifier gain (top panel) and zero-delayed second-order correlations $g^{(2)}_{aa}(0)$ (red circles), $g^{(2)}_{bb}(0)$ (blue circles), and $g^{(2)}_{ab}(0)$ (black triangles) with different pump power (bottom panel). The inserted figure in the bottom panel is the time-delayed correlation corresponding to the pump power of $-102.36$ dBm. The grey dashed line represents the theoretical value of the auto-correlation, and the black dashed line is the predicted cross-correlation from the parametric amplifier gain. (b) Time-delayed second-order cross-correlations at different pump powers. The dashed lines are the theoretical fitting of the cross-correlation profile.} 
    \label{fig:g2result}
\end{figure}

The second-order auto- and cross-correlation are calculated simultaneously in real-time. The results for each drive power are averaged over 500 computational buffers, with $2^{21}$ samples per buffer. To estimate the error bars, the 500 buffers are divided into 10 segments of equal numbers, and the standard deviation is computed from the results of these segments. The second-order cross-correlation result is shown in Fig. \ref{fig:g2result} (a). By sweeping the pump power from $-100.82$ dBm to $-102.79$ dBm, we observe an increase on $g^{(2)}_{ab}(0)$ from $2.21\pm0.01$ to $11.91\pm0.68$. This shows a violation of the classical relation shown in Eq. (\ref{equ:inequality}) and thus confirms the generation and non-classicality of the photon pairs. The increasing $g^{(2)}_{ab}(0)$ shows higher photon-pair purity with reduced squeezing parameter in Eq. (\ref{equ:tms_sq_fock_basis}). The inserted figure is the time-delayed second-order correlation corresponding to $P_{\mathrm{drive}} = -102.36$ dBm. The auto-correlation remains consistent at around 2 over the power sweep. With a lower drive ($P_{\mathrm{drive}} < -102.79$ dBm), it is hard to resolve auto-correlation with a low error bar due to strong thermal-noise correlation and limitations of the measurement setup (See Appendix. \ref{sec:dataprocessing} for detail). The cross correlation $g^{(2)}_{ab}$, on the other hand, is less susceptible to the background noise as the noise from the two modes is not correlated. 

The parametric amplifier gain $G_{\mathrm{NKPA}}$ is connected to the mean photon number in each itinerant mode as $\langle a^\dagger a\rangle = \langle b^\dagger b\rangle = G_{\mathrm{NKPA}}-1$. The zero-delayed second-order cross-correlating can be expressed as 
\begin{equation}
        g^{(2)}_{ab} (0)  = 2 + \frac{1}{2\langle a^\dagger(t) a(t) \rangle}
                      = 2 + \frac{\eta}{2(G_{\mathrm{NKPA}}-1)}.
\end{equation}
In this context, we have introduced a unitless parameter $\eta$, which ranges from 0 to 1, to compensate for the imperfections observed in the resulting two-mode states. These imperfections include non-zero intracavity photon occupancy and imperfect symmetry between the two modes. In an ideal situation, $\eta=1$. The definition of $\eta$ is similar to the noise-reduction-factor (NRF) defined in the reference \cite{westig2017emission}. We fit $g^{(2)}_{ab} (0)$ with corresponding gain, as shown in Fig. \ref{fig:g2result} (a), and get $\eta=0.72$. 

In Fig. \ref{fig:g2result} (b), we show the time-delayed cross-correlation $g^{(2)}_{ab}(\tau)$ with pump power at $-102.79$ dBm, $-102.36$ dBm, and $-101.95$ dBm. The dashed line shows a fitting of the decay equation $g^{(2)}_{ab}(\tau) = 1 + \gamma_c/(2R) \cdot e^{-\gamma_c|\tau|}$, where $\gamma_c$ is the characteristic decay rate and $R$ is the detected photon-pair rate \cite{fortsch2013versatile}. At a pump power of $-102.79$ dBm, the fitted $\gamma_c$ is $5.90$ MHz, and $R$ is $0.21$ MHz. Employing a similar approach, we observe the photon-pair rates at pump powers of $-102.36$ dBm and $-101.95$ dBm to be $0.52$ MHz and $0.99$ MHz, respectively.

\begin{figure}[t]
    \centering
    \includegraphics[width=\linewidth]{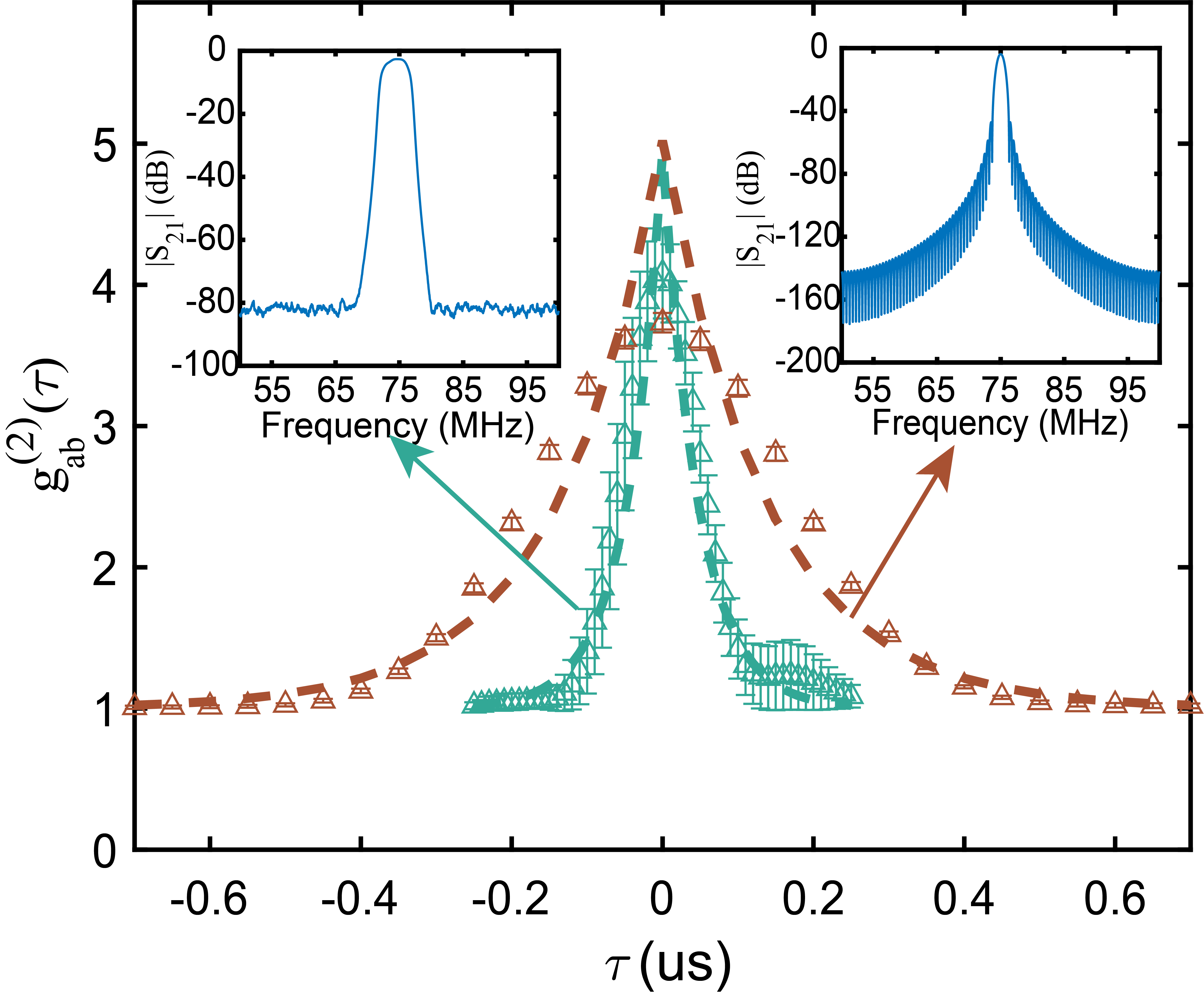}
    \caption{The illustration compares $g^{(2)}_{\mathrm{ab}}(\tau)$ under different measurement bandwidths as determined by the physical filter (represented in brown) and the digital filter (depicted in cyan). The inserts showcase the filter response function within the frequency domain. The left section visualizes the physical bandpass filter, which has a bandwidth of 3.88 MHz, while the right section portrays the digital bandpass filter, with a bandwidth of 0.86 MHz.}
    \label{fig:bandpassFilterCompare}
\end{figure}

The measurement bandwidth, which is set by the physical and digital filters used to improve the SNR, constrains the detectable rates of photon-pair occurrences. To demonstrate this, we adjust the measurement bandwidth by either incorporating or excluding the narrower digital bandpass filter, which is executed on the GPU. In both scenarios, the physical bandpass filter is kept in place, and therefore, the measurement bandwidth is set to be 3.88 MHz without the digital filter and 0.88 MHz with the digital filter. The results of the second-order cross-correlation are depicted in Fig. \ref{fig:bandpassFilterCompare}. The fitting for both cases yields a very consistent ratio of $\gamma_c/(2R)$, with respective values of $3.94$ and $3.99$. However, the characteristic decay rate, $\gamma_c$, varies with $20.94$ MHz and $7.026$ MHz, respectively, indicating photon-pair rates of $2.66$ MHz and $0.88$ MHz. These ratios closely correspond to their relative measurement bandwidths. 

\begin{figure}[t]
     \centering
     \includegraphics[width=\linewidth]{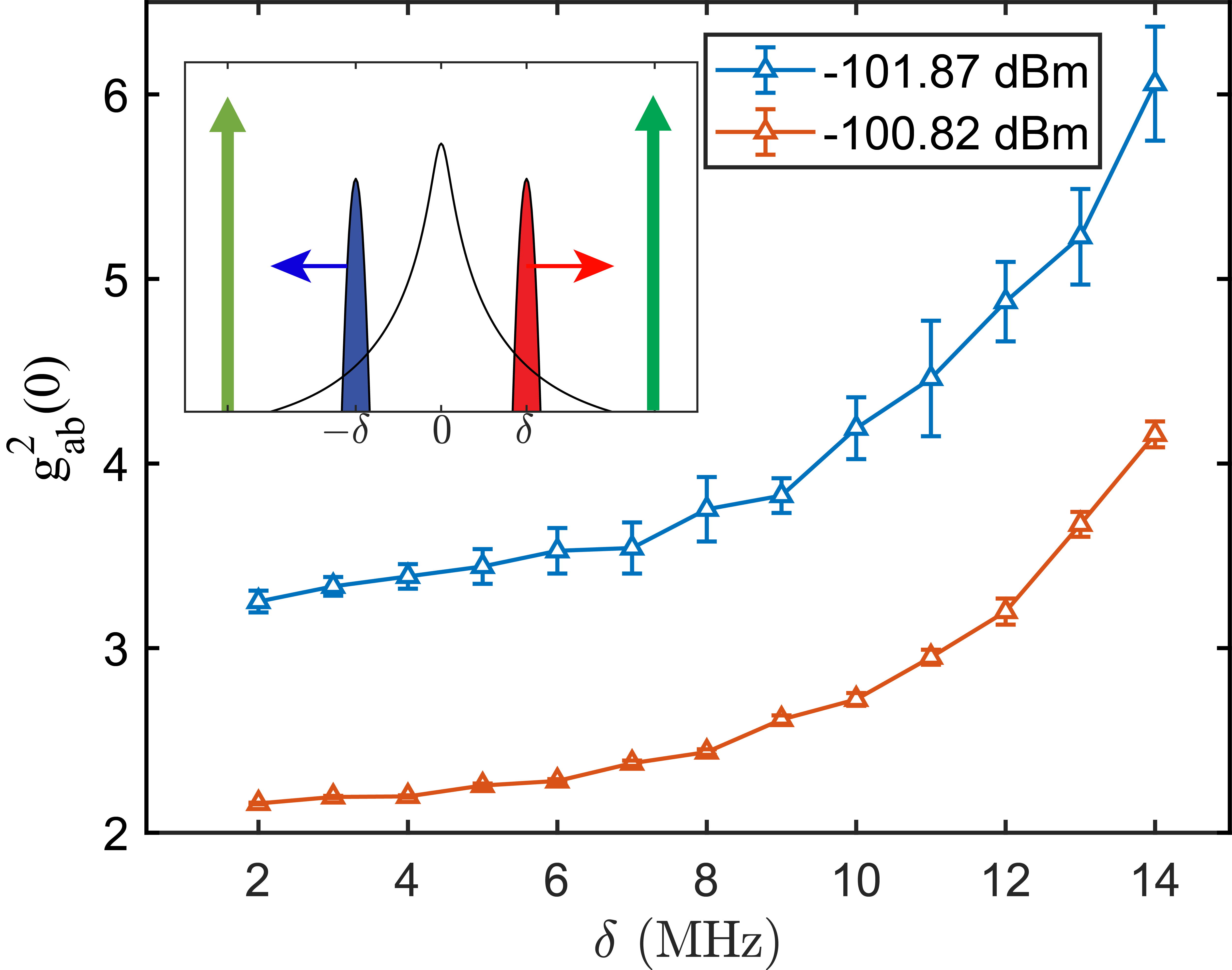}
     \caption{Zero-time delayed second-order cross-correlation corresponding to different demodulation detuning for pump power at $-101.87$ dBm and $-100.82$ dBm. The inserted illustrate the change of demodulation detuning for the two modes.}
     \label{fig:bandwidth}
 \end{figure}

The two modes can be demodulated at different frequencies, contingent upon the device's bandwidth. The results presented in Fig. \ref{fig:g2result} are based on a $2$ MHz detuning. As depicted in Fig. \ref{fig:bandwidth}, we increment the detuning from 2 MHz to 14 MHz with a fixed pump power of -101.97 dBm and -100.82 dBm, and subsequently measure the cross-correlation between the two modes. The zero-delayed correlation augments due to the reduction in the squeezing parameter. Remarkably, non-classicality is preserved up to 14 MHz, signifying that two modes separated by 28 MHz still retain non-classical behavior.  The bandwidth of the source is constrained by the gain-bandwidth product, which in our case is 59 MHz. 

In conclusion, we have demonstrated a novel kinetic-inductance device capable of generating non-classical microwave photon pairs. In comparison to Josephson junction (JJ)-based devices, this innovative design offers easier fabrication and eliminates the need for magnetic field shielding when installed within a cryogenic environment. Furthermore, the device's high transition temperature facilitates radiative cooling \cite{xu2020radiative, xu2023radiatively}. As a result, this technology promises significant reductions in both space requirements and heat budget when installed in a dilution refrigerator.

\begin{acknowledgments}
The authors would like to thank Dr. Yong Sun, Dr. Lauren McCabe, Mr. Kelly Woods, and Dr. Michael Rooks for assistance in device fabrication. We acknowledge funding support from DOE Office of Science, National Quantum Information Science Research Centers, Co-design Center for Quantum Advantage (C2QA), Contract No. DE-SC0012704. We acknowledge partial funding support from the Office of Naval Research on the development of nitride-based superconductors (under Grant No. N00014-20-1-2126).\end{acknowledgments}

\appendix
\section{Hamiltonian derivation}
\label{app:Hamiltonian}
The Hamiltonian of the NKPA can be expressed in the interaction picture as
\begin{equation}
\begin{split}
    \frac{H_{\mathrm{int}}}{\hbar} & = -\frac{1}{4\hbar} \frac{L_{k0}}{I^{*2}} I_{ZPF}^4 (a+a^{\dagger})^4\\
    & \approx  -\frac{6}{4\hbar} \frac{L_{k0}}{I^{*2}} I_{ZPF}^4 (2a^\dagger a + a^{\dagger 2}a^2),
    \label{equ:hamiltonian}
\end{split}
\end{equation}
where $a$ is the bosonic operator, $L_{k0}$ is the kinetic inductance of the nanowire, $I_{ZPF}$ is the zero-point current of the resonator, and $I^*$ is the characteristic current of the nanowire. The approximation is applied by dropping the fast-rotating term and constant \cite{parker2022degenerate}. We consider two symmetrical detuned pumps as classical drive, and assume the two correlated modes as $\delta a$ and $\delta b$. The NKPA intracavity field $a$ can be expressed in the rotating frame of the resonance frequency as

\begin{equation}
    a = A_1 e^{i (\Delta t + \phi_1)} + A_2 e^{i (-\Delta t + \phi_2)} + \delta a e^{i \delta t} + \delta b e^{-i \delta t},
    \label{equ:modeExpression}
\end{equation}
where $A_{1, 2}$, $\phi_{1,2}$ are the pump amplitudes and phases. Plugging Eq. (\ref{equ:modeExpression}) into Eq. (\ref{equ:hamiltonian}), we derive the following Hamiltonian after rotating-wave approximation 

\begin{equation}
\begin{split}
        \frac{H_{\mathrm{int}}}{\hbar} = & - \frac{K}{4}\left(\delta a^2 \delta a^{\dagger2} + \delta b^2 \delta b^{\dagger2} \right) \\
        & - K \delta a \delta a^\dagger \delta b \delta b^\dagger \\
        & + K\left(A_1^2 + A_2^2\right) \left(\delta a^\dagger \delta a + \delta b^\dagger \delta b \right) \\
        & + \left(\epsilon\delta a^\dagger \delta b^\dagger + \epsilon^* \delta a \delta b \right),
\end{split}
    \label{equ:hamiltonian_final}
\end{equation}
where the vacuum Kerr nonlinearity $K$ and interaction strength $\epsilon$ are defined as 
\begin{equation}
\begin{split}
        K & = \frac{6}{\hbar} \frac{L_{k0}}{I^{*2}}I_{ZPF}^4, \\
        \epsilon & = K A_1 A_2 e^{i(\phi_1 + \phi_2)}.
\end{split}
\end{equation}
The first term in Eq. (\ref{equ:hamiltonian_final}) is due to the self-Kerr effect and is negligible when $\delta a$, $\delta b$ are weak signal. The second and third term refere to the cross-Kerr effect, and the last term represents non-degenerate parametric down-conversion. 

To relate the pump power $P$ and the amplifier gain $G_{\mathrm{NKPA}}$, we consider the pump detuning $\Delta$, we can write the interaction strength as $\epsilon = K P\kappa_{\mathrm{ex}}^2/(\hbar \omega_s(\Delta^2+\kappa^2))$, where $\kappa_{\mathrm{ex}}$ is the external loss rate, and $\kappa = \kappa_{\mathrm{ex}} + \kappa_{\mathrm{in}}$ is the total loss rate. The squeezing coefficient $r = \epsilon \tau$, where $\tau=1/\kappa$ is the cavity decay time. Therefore, we can derive
\begin{equation}
    G_{\mathrm{NKPA}} = (\cosh\left[K\frac{P}{\hbar \omega_s} \frac{\kappa_{\mathrm{ex}}^2}{\kappa(\Delta^2 + \kappa^2)}\right])^2.
\end{equation}

\section{Cryogenic setup and control instrument}
\label{sec:cryo_setup_control}

\begin{figure}
    \centering
    \includegraphics[width=\linewidth]{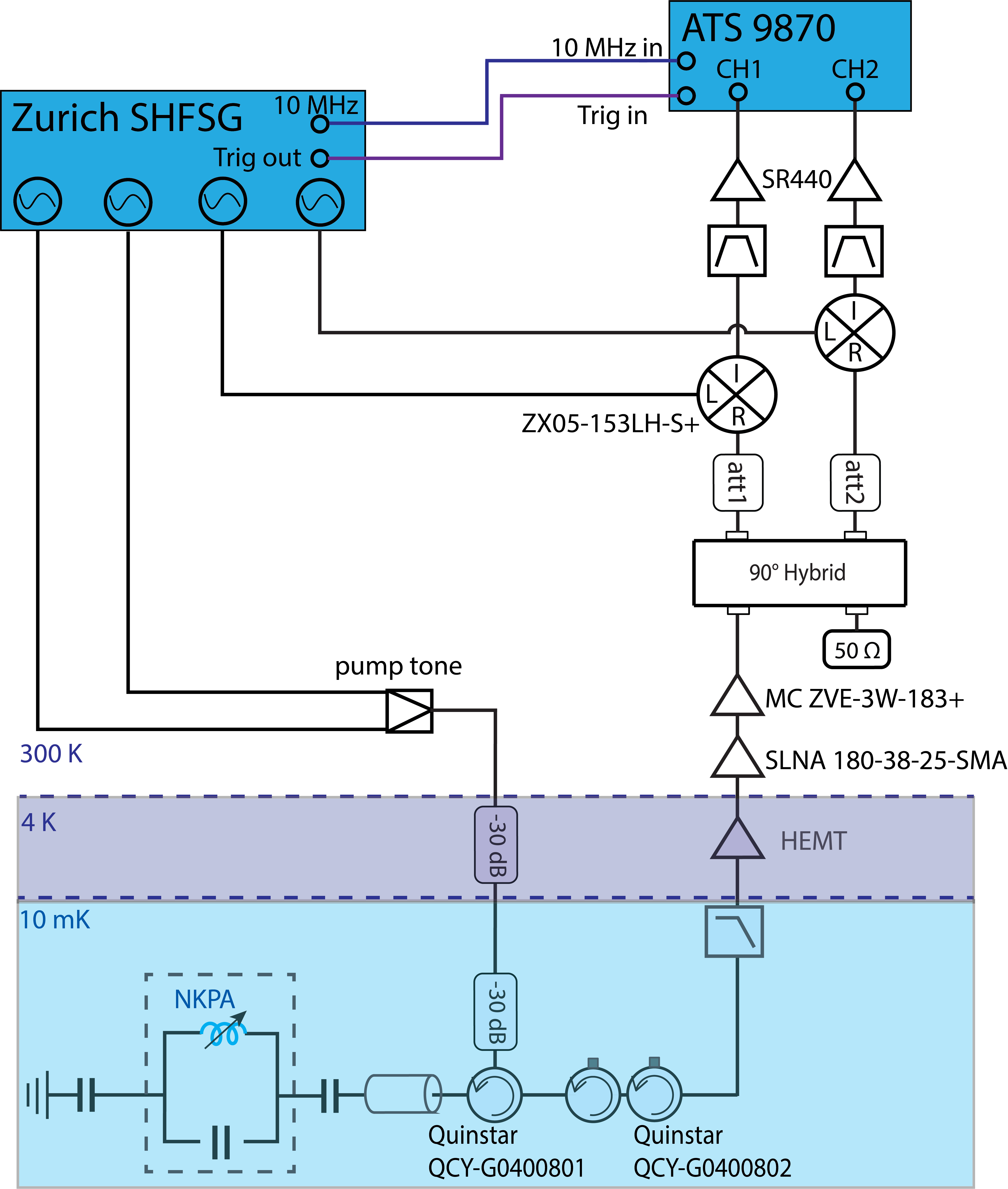}
    \caption{Schematic representation of the cryogenic wiring along with the room-temperature control and measurement arrangement. A Quinstar double junction circulator (QCY-G0400802) is employed to offer an isolation of 40 dB from the output line. The elements labeled as ``att1" and ``att2" are attenuators, utilized to prevent power saturation in the subsequent mixers.}
    \label{fig:detailed_setup}
\end{figure}

The detailed schematics of the cryogenic wiring along with the room-temperature measurement setup are depicted in Fig. \ref{fig:detailed_setup}. The two pump signals are generated by a Zurich Instrument SHFSG signal generator, combined, and attenuated by a total of 60 dB before being fed into the NKPA device. The emitted two-mode radiation, in conjunction with the residual pump, is reflected and channeled to the output line through a single junction circulator (QCY-G0400802). A double junction circulator (QCY-G0400802) is employed to further reject any reflected tone from the output line. The microwave radiation output is then amplified by a HEMT (Low Noise Factory, LNF-LNC$\_$16B) situated at the 4K plate and transferred to the room-temperature setup.

Within the room-temperature configuration, the signal is initially amplified using a low noise amplifier (SLNA 180-38-25-SMA), followed by a high power amplifier (Mini-circuit ZVE-3W-183+). Subsequently, the signal is split via a $90^\circ$ hybrid (RF-Lambda RFHB02G18GVT). It should be noted that up until this point, the output microwave still retains the components of the two pump tones. As such, an attenuator is connected to each line to reduce the overall power, thus preventing the mixer (Mini-Circuit ZX05-153LH-S+, 1 dB compression power at 2 dBm) from reaching saturation. The local oscillators for the two mixers are sourced from the SHFSG and are phase coherent with the pump tones. The signal is down-converted to 75 MHz, which is well-separated from the down-converted pump tone, enabling efficient filtering by a narrow bandpass filter (Texscan Microwave Products 6BL75.1/3.0-EE). The filtered signal is further amplified (using SR400) and dispatched to the digitizer (AlazarTech ATS9870).

The 8-bit digitizer operates at 100 MSp/s with a $\pm 40$ mV input range. To commence data acquisition, the SHFSG sends TTL triggers to the digitizer at a frequency of 5 Hz. Each received trigger initiates the acquisition of a record consisting of $2^{21}$ data points. Two such records constitute a data buffer sent to the GPU for processing. These two records correspond to two triggers, where the first trigger has the SHFSG sourcing the two-pump tone, and the second trigger shifts one of the pump tones by 80 MHz, negating gain on the NKPA. These records correspond to the ``ON" and ``OFF" data states. This data buffer is subsequently transferred to a GPU (NVIDIA GeForce RTX 2080 Ti) for correlation calculations.

\section{Data Processing}
\label{sec:dataprocessing}

The signal is demodulated to $75$ MHz before being sent to the digitizer, which operates at a sampling rate of $100$ MSp/s. As a result, the signal frequency is aliased to $25$ MHz. It is important to note that this aliasing process does not decrease the signal-to-noise ratio since the narrow bandpass filter has already filtered out the noise at frequencies higher than $100$ MHz. Therefore, the signal-to-noise ratio is preserved in the digitized signal at $25$ MHz.

The second-order cross-correlation is calculated with Eq. (\ref{equ:g2_cross})
\begin{equation}
\label{equ:g2cross_expand}
    \begin{split}
        g^{(2)}_{ab}(\tau) & = \frac{\langle b^{\dagger} (0)a^{\dagger}(\tau)a(\tau)b(0) \rangle )}{\langle a^{\dagger}(\tau) a(\tau) \rangle \langle b^{\dagger}(0) b(0) \rangle}\\
        & = \frac{\langle n_{a}^s(\tau) n_{b}^s(0) \rangle}{\langle n_{a}^s(0) \rangle \langle n_{b}^s(0) \rangle} \\
        &  = \frac{\langle n_{a}(\tau) n_{b}(0) \rangle - \langle n_{a}(\tau) \rangle \langle n_{b}^n(0)  \rangle }{\langle n_{a}^s(0) \rangle \langle n_{b}^s(0) \rangle} \\
        & + \frac{- \langle n_{a}^n(\tau)\rangle \langle n_{b}(0)  \rangle + \langle n_{a}^n(\tau)\rangle \langle n_{b}^n(0)  \rangle}{\langle n_{a}^s(0) \rangle \langle n_{b}^s(0) \rangle},
    \end{split}
\end{equation}
where $n^n(\tau)$ and $n(\tau)$ represent the output field first-order autocorrelation of the ``OFF" and ``ON" process, and $n^s(\tau) = n(\tau) - n^n(\tau)$ is first-order autocorrelation of the output signal field. It is important to note that we assume no correlation between the noise terms in the two modes, enabling us to express the mean of the product of two variables as the product of their means.

To calculate the second-order auto-correlation, it is no longer valid to implement the calculations similar to Eq. \ref{equ:g2cross_expand} due to the emergence of non-commuting terms. To illustrate, we write the field operator as $S_a(t) = \sqrt{G_{\mathrm{line}}}(a(t) + h_a^{\dagger}(t))$, where $G_{\mathrm{line}}$ is the gain of the output line, $\langle S_a^{\dagger}(t)S_a(0) \rangle = n_a(t)$, $G_{\mathrm{line}}\langle a^{\dagger}(t)a(0) \rangle = n^s_a(t)$, and $G_{\mathrm{line}} \langle h_a^{\dagger}(t)h_a(0) \rangle = n^n_a(t)$. The noise term $h_a^{\dagger}(t))$ is determined by the HEMT. Therefore, the first order correlation 
\begin{equation}
\begin{split}
        n_a^s(t) & = G_{\mathrm{line}} \langle a^{\dagger}(t)a(0) \rangle \\
        & =  \langle S_a^{\dagger}(t)S_a(0) \rangle - G_{\mathrm{line}}\langle h^{\dagger}_a(t)h_a(0) \rangle - \delta(t) \\
        & = n_a(t) - n_a^n(t) - \delta(t),
\end{split}
\end{equation}
where $\delta_a$ comes from the commutation relation $\left[h_a(t), h^{\dagger}_a(0) \right] = \delta(t)$. The $\delta$ function can be eliminated by the limited measurement bandwidth \cite{da2010schemes, grimm2015josephson}. The second-order correlation can be calculated in a similar fashion by rearranging the field operators with the commutation relation \cite{grimm2015josephson}
\begin{equation}
    \begin{split}
        \langle n_{a}(\tau) n_{a}(0) \rangle & = \langle S_a^\dagger(0) S_a^\dagger(\tau) S_a(\tau) S_a(0) \rangle \\
        & = \langle n_{a}^s(\tau) n_{a}^s(0) \rangle + \langle n_{a}^s(\tau) n_{a}^n(0) \rangle \\
        & + \langle n_{a}^n(\tau)n_{a}^s(0) \rangle + \langle n_{a}^n(\tau) n_{a}^n(0) \rangle \\
        & + 2\langle n_{a}^n(0)n_{a}^s(0) \rangle. \\
    \end{split}
\end{equation}
This gives
\begin{equation}
\label{equ:g2auto}
\begin{split}
g^{(2)}_{aa}(\tau) & = \frac{\langle n_{a}^s(\tau) n_{a}^s(0) \rangle}{\langle n_{a}^s(0) \rangle \langle n_{a}^s(0) \rangle}\\
& = \frac{\langle n_{a}(\tau) n_{a}(0) \rangle - \langle n_{a}^s(\tau) \rangle \langle n_{a}^n(0) \rangle }{\langle n_{a}^s(0) \rangle \langle n_{a}^s(0) \rangle} \\
& + \frac{- \langle n_{a}^n(\tau) \rangle \langle n_{a}^s(0) \rangle - \langle n_{a}^n(\tau) n_{a}^n(0) \rangle}{\langle n_{a}^s(0) \rangle \langle n_{a}^s(0) \rangle} \\
& + \frac{- 2 \langle n_{a}^n(0) \rangle \langle n_{a}(0) \rangle }{\langle n_{a}^s(0) \rangle \langle n_{a}^s(0)\rangle},
\end{split}
\end{equation}
where we have dropped the terms with the $\delta$ function. The expression for $g^{(2)}_{bb}(\tau)$ can be derived similarly. In practice, we replace field $a$ and $b$ with their corresponding complex envelop operator $S_{a, b}(t) = X_{a, b}(t) + iP_{a, b}(t),$ where $X_{a, b}$ and $P_{a, b}$ are the two quadratures of the complex envelope. To establish the correspondence between the digitized voltage and the quadratures, we can rewrite the voltage as
\begin{equation}
    V(t) = X(t) \cos(2\pi f t) + P(t) \sin(2\pi f t),
\end{equation}
with $f = 75$ MHz is the carrying frequency. With a sampling rate of $100$ MSp/s, the sampled voltage can be rewritten as
\begin{equation}
    V_n = V(n \Delta T)
    \begin{cases}
      X_n (-1)^\frac{n}{2} & \text{if $n$ is even}\\
      P_n (-1)^\frac{n+1}{2}& \text{if $n$ is odd}\\
    \end{cases}
    \label{equ:digitizing}
\end{equation}
where $\Delta T =$ 10 ns is the time interval between two samples. Eq. (\ref{equ:digitizing}) shows that the sample voltages are the two quadratures with alternating signs. If we manually correct the sign, we are effectively demodulate the signal to zero frequency. This process is known as frequency translation in digital signal processing \cite{lyons1997understanding}.

In order to proficiently filter $1/f$ noise, a Finite Impulse Response (FIR) bandpass filter, incorporating a Hann window, has been utilized. In the frequency domain, the filter's frequency response, $H(\omega)$, is multiplied with the signal spectrum. Concurrently, in the time domain, the signal is convolved with $h(n) = \mathcal{F}^{-1} { H(\omega) }$, where the order, $N$, is $200$. The frequency response can be viewed in Fig. \ref{fig:bandpassFilterCompare} (inserted on the left) and the bandwidth is 0.86 MHz. 

The ability to measure the correlation at a smaller power (than -102.79 dBm) is limited by two factors. Firstly, the quantization noise emanating from the analog-to-digital conversion (ADC) imposes a fundamental limitation. Our digitizer incorporates an 8-bit ADC, enabling the translation of analog signals into integers ranging from -127 to 128. This conversion process inevitably leads to quantization errors. The quantization error can be quantified as quantization signal-to-noise ratio (SNR) and is given by $\mathrm{SNR} \approx 6.08\times Q + 1.76$ dB, where Q is 8 for 8 bits ADC \cite{kester2009taking}. In other word, this set a lower bound of the noise power due to the quantization, defined by the signal power that takes the full dynamic range of the ADC divided by the quantization SNR. This means for signal power smaller than the quantization noise power would be challenging to measure. The second factor is the emergence of the delta terms. As discussed in earlier, due to the commutation relation, there will be $\delta(t)$ terms in the second-order auto-correlation. The $\delta(t)$ terms would be small with a finite bandwidth. However, if the signal component is small, the contribution from $\delta(t)$ terms cannot be neglected. This will make the measurement very unstable and not accurate \cite{grimm2015josephson}. In conclusion, we are not able to decrease the pump power as the result becomes very volatile and not accurate, due to the limitation of the measurement setups.

\section{Non-classicality analysis}
In this section, we give a detailed analysis on the relation between the parametric amplifier gain and the second-order cross-correlation. The input and output relation of the interactivity mode can be expressed as 
\begin{equation}
    \begin{split}
        a & = \sqrt{G_{\mathrm{NKPA}}} a_{\mathrm{in}} + \sqrt{G_{\mathrm{NKPA}}-1} b_{\mathrm{in}}^{\dagger}, \\
        b & = \sqrt{G_{\mathrm{NKPA}}} b_{\mathrm{in}} + \sqrt{G_{\mathrm{NKPA}}-1} a_{\mathrm{in}}^{\dagger}, \\
    \end{split}
\end{equation}
where $G_{\mathrm{NKPA}} = (\cosh{r})^2$ is the parametric gain of NKPA, $a_{\mathrm{in}}$ and $b_{\mathrm{in}}$ are the intracavity modes. The mean photon number of the output two-mode squeezed state is 
\begin{equation}
\begin{split}
    \langle a^\dagger a\rangle  & = G_{\mathrm{NKPA}} \langle a_{\mathrm{in}}^\dagger a_{\mathrm{in}}\rangle + (G_{\mathrm{NKPA}}-1)\langle b_{\mathrm{in}} b_{\mathrm{in}}^\dagger\rangle \\
    \langle b^\dagger b\rangle  & = G_{\mathrm{NKPA}} \langle b_{\mathrm{in}}^\dagger b_{\mathrm{in}}\rangle + (G_{\mathrm{NKPA}}-1)\langle a_{\mathrm{in}} a_{\mathrm{in}}^\dagger\rangle \\
\label{equ:output_photon_number}
\end{split}
\end{equation}
Considering intracavity photon occupancy $\langle a_{\mathrm{in}}^\dagger a_{\mathrm{in}} \rangle = \langle b_{\mathrm{in}}^\dagger b_{\mathrm{in}} \rangle = n$, then the above expression can be simplified as
\begin{equation}
\begin{split}
    \langle a^\dagger a\rangle = \langle b^\dagger b\rangle & = 2(G_{\mathrm{NKPA}}-1)n + G_{\mathrm{NKPA}}-1 \\
    & = (G_{\mathrm{NKPA}}-1)/\eta,
\end{split}
\end{equation}
where $\eta = 1/(1+2n)$, and $\eta = 1$ when both modes are at the ground state. Here we consider that all the imperfection come from the non-zero intracavity photon occupancy. The zero-delayed second-order correlation is linked to the intracavity photon number as established by \cite{westig2017emission}
\begin{equation}
    g^{(2)}_{ab} (0) = 2 + \frac{\eta}{2(G_{\mathrm{NKPA}}-1)}.
\label{equ:appendix_g2}
\end{equation}
In the main text, the fitting result gives $\eta=0.72$. That means if all the imperfections are from the non-zero intracavity photon occupancy, then $n=(1/\eta - 1)/2 = 0.19$. The unitless parameter $\eta$ can also be connected with noise-reduction factor defined in reference \cite{westig2017emission} with relation $\textrm{NRF} = 1 - \eta / 2$. In our case, this yield $\textrm{NRF}=0.64$. 

\bibliography{apssamp}

\end{document}